\documentclass[preprint,12pt]{elsarticle}

\usepackage{amsmath,amssymb,amsfonts}

\usepackage{xcolor}
\usepackage{booktabs}
\usepackage[linesnumbered,ruled,vlined]{algorithm2e}
\usepackage{array}
\usepackage{multirow}
\usepackage{rotating}

\journal{arXiv}

\begin{document}

\begin{frontmatter}



\title{Guarding the Grid: Enhancing Resilience in Automated Residential Demand Response Against False Data Injection Attacks}


\author[inst1]{Thusitha Dayaratne}
\author[inst1]{Carsten Rudolph}
\author[inst1]{Ariel Liebman}
\author[inst1]{Mahsa Salehi}

\affiliation[inst1]{organization={Faculty of IT, Monash University},
            city={Clayton},
            postcode={3800}, 
            state={Victoria},
            country={Australia}}

\begin{abstract}
Utility companies are increasingly leveraging residential demand flexibility and the proliferation of smart/IoT devices to enhance the effectiveness of residential demand response (DR) programs through automated device scheduling. However, the adoption of distributed architectures in these systems exposes them to the risk of false data injection attacks (FDIAs), where adversaries can manipulate decision-making processes by injecting false data. Given the limited control utility companies have over these distributed systems and data, the need for reliable implementations to enhance the resilience of residential DR schemes against FDIAs is paramount. In this work, we present a comprehensive framework that combines DR optimisation, anomaly detection, and strategies for mitigating the impacts of attacks to create a resilient and automated device scheduling system. To validate the robustness of our framework against FDIAs, we performed an evaluation using real-world data sets, highlighting its effectiveness in securing residential DR systems.

\end{abstract}




\end{frontmatter}


\section{Introduction}
Distributed demand response (DR) optimisation methods \cite{parvania2010demand,nan2018optimal} are gaining momentum with the emergence of the Internet of things (IoT) technology and advancements in the smart grid. These schemes calculate optimal demand distributions in coordination by exchanging price signals and demand forecasts between home energy management systems (HEMS)/home automation processes and demand coordinators. The use of day-ahead pricing with automated device scheduling minimises manual interactions on these schemes \cite{He2018}. Utilities are also leveraging the residential demand forecast with the introduction of transactive energy markets \cite{daneshvar2019transactive} to determine market behaviours that could influence the economic dispatch processes in power grids. Despite having significant advantages of residential demand forecasts, utility companies cannot access individual forecasts as consumers possess private cost/welfare functions that they optimise. Thus, utility companies require to use aggregated demand/load forecasts of communities instead of granular individual forecasts. These restrictions and distributed architectures allow malicious individuals and organised groups to execute cyberattacks to gain financial advantages or induce malevolent impact. In particular, false data injection attacks (FDIAs) \cite{Liu2009} where adversaries inject or alter the data can trigger incorrect or sub-optimal power flows, excessive or insufficient generation, financial losses, and inconvenience for the users or even physical damages depending on the severity of attacks \cite{Pilz2019,Barreto2019}. 

FDIAs are imminent with the increasing use of DR systems \cite{barreto2020attacking,Dayaratne2019}. 
Many works have explored FDIAs on power grid state estimation processes and bulk energy markets. However, the probability of such FDIAs is significantly lower, as it requires significant compromises on well-secured grid infrastructure and complete knowledge of configurations and architectural information. Thus, adversaries strive to mount FDIAs from components/devices that reside on the consumer side (outside the control of utility companies), such as smart meters, HEMS, and smart appliances \cite{Soltan2018,Huang2019}. These attacks can result in a similar or much higher impact compared to traditional FDIAs as operators utilise consumer forecasts in generating optimal solutions. Thus, proper security analyses and secure implementations are essential to \cite{Pilz2019,Barreto2019,Dayaratne2019} the realisation of reliable DR schemes and energy markets.

Limited research has focused on analysing and enhancing the resilience of large-scale, residential, device scheduling-based DR systems against novel FDIAs as they are in inception. Thus, in this work, first, we formalise the distributed residential DR and the impact of FDIAs on the DR context using a commonly adopted scheme. Then we implement a resilient, automated distributed device scheduling DR framework that can withstand FDIAs. Though we utilise a distributed DR framework in our implementation, the framework is not limited to distributed DR schemes. The framework can be used with centralised and decentralised DR schemes that leverage residential demand forecasts. In particular, we make the following contributions:

\begin{enumerate}
    \item Leverage DR optimisation, anomaly detection and impact mitigation methods into an integrated solution for resilient distributed device scheduling.
    \item Depict a practical implementation of the resilient distributed device scheduling framework that integrates the different components.
    \item Evaluate the implemented framework using a real-world dataset and discuss limitations.
\end{enumerate}

The paper is organised as follows: Section 2 provides background and discusses the current state of FDIA in DR schemes and the smart grid. Section 3 presents the system model and FDIA. In Section 4, we introduce the resilient device scheduling framework and its modules. The implementation details and experimental results are presented in Sections 5 and 6, respectively. Section 7 discusses the resiliency of the proposed framework and potential extensions. The paper concludes in Section 9.

\section{Related Work}
Motivated by the initial work of Liu et al. \cite{Liu2009}, a considerable amount of literature has been published on FDIAs. Most of the literature is focused either on FDIAs against the state estimation or energy dispatching scenarios such as SCED~\cite{Liang2017}. However, the plausibility of such attacks is low, as it is arduous to compromise highly protected critical infrastructure unless adversaries possess excessive budgets, information, and other adequate competencies. In contrast, customer-owned edge devices such as smart appliances, smart meters, and HEMSs in smart energy networks enable much broader attack surfaces. Nevertheless, the impact of FDIAs on residential DR schemes using such edge devices and potential countermeasures are not sufficiently explored.

Srikantha and Kundur introduced a resilient distributed real-time DR strategy using population games~\cite{Srikantha2017}. Compared to this work, they used DR agents with pre-agreed levels of demand reduction, whereas we focused on demand shifting without any curtailment. Further, our focus is independent of a particular DR algorithm. Liu et al. used a partially observable Markov decision process to help utility companies detect and fix coordinated attacks on smart meters \cite{Liu2017}. In contrast, our approach avoids manual inspections and associated costs at the HEMS/smart meter level. Moreover, they rely on the real consumption values instead of forecasted demand. Anuebunwa et al. proposed the use of a genetic algorithm to optimise forecasts using pricing and occupancy data, when forecasts are affected by noise \cite{Anuebunwa2017,Anuebunwa2018}. However, it does not address the impact when communication channels are used to inject the false data and the computation and resource overhead of running optimisation at the HEMS, whereas in our work, there is no overhead for the HEMSs. Barreto and Cardenas compared the benefits obtained by adversaries using FDIAs in a decentralised DR scheme versus a Direct Load Control system. Although they designed a penalty scheme against FDIAs, the scheme could not fully compensate victims of attacks \cite{Barreto2019}. Sethi et al. proposed a resilient scheduling algorithm against FDIAs. However, their approach assumes that the utility has complete information about individual house schedules, raising privacy concerns~\cite{Sethi2020}.

Existing literature focuses on state estimation or specific aspects of DR algorithms, such as impact, detection, and countermeasures. However, there is a gap in analysing the end-to-end flow of a residential device scheduling framework against FDIAs. This work fills that gap by examining the complete flow of a resilient residential device scheduling-based DR scheme under FDIAs. Our implementation is versatile and can be applied to any device scheduling DR scheme as long as the scheme follows a similar flow, regardless of the specific optimisation algorithm used.

\section{Preliminaries}
\label{sec:prel}

\subsection{DR Model}
We have adopted a two-stage pricing model in our residential DR scheme that aligns with the wholesale energy market structure. In particular, day-ahead pricing is used to schedule appliances \cite{He2018,LiChenLow2011,Mhanna2016}, and real-time pricing is used for billing purposes. We assume that most users will follow their forecasts. Thus, the day-ahead price signal is about approximately to the real-time price signal.

A day is divided into $P$ equal pricing slots, where each pricing slot comprises $k(>1)$ scheduling slots. Before a day starts, the utility company/demand coordinator sends the day-ahead price signal for the following day to all smart meters. The price signal is a vector of $P$ price values. Each value corresponds to the unit price of a pricing slot. Unit price values are indexed from $1$ to $p$. Each house has several controllable devices that need to be scheduled. The house owner configures the earliest start time, the latest finish time, preferred start time, device demand, running duration, and the penalty factor for each device. A device can operate within one or more scheduling slots. The penalty factor is used to calculate the user inconvenience when a device does not start at its preferred start time. The HEMS locally schedules devices between the configured earliest start time and the latest finish time of individual devices, based on the received day-ahead price signal to achieve a specific optimisation target (lower cost, lower inconvenience, etc.). Upon scheduling appliances, the forecasted demand profiles of individual households are sent to an aggregator, which accumulates all forecasts and forwards the cumulative demand forecast to the utility company. 
End-to-end, efficient encryption and privacy-preserving aggregation methods can be applied to ensure the privacy of forecasts \cite{lu2012eppa,he2017efficient,abdallah2016lightweight} as individual demand forecasts are not required by utilities. They only require consumption data for billing purposes. Thus, depending on the protocols used, even the aggregator will not possess access to individual demand forecasts.
The utility company adjusts the price signal based on the received forecast to achieve its optimisation targets and resends the adjusted price signal back to SMs/HEMSs. The process continues until the system converges to the optimal solution that minimises the total cost of the community. 

\subsection{Optimisation}
For simplicity, we restrict the explanation to the following case.
\begin{enumerate}
\item We identify two pricing slots $p_i$ and $p_j$ where $i\neq j$. 
\item The appliances run on either one of the two pricing slots or in both pricing slots.
\item The unit price of a slot is a function of the demand of the same slot, where the function is monotonically increasing.
\end{enumerate}

We denote $ad_{i(init)}$, $ad_{i(opt)}$, $ad_{j(init)}$ and $ad_{j(opt)}$
as initial and optimised aggregated demand values of the i\textsuperscript{th} and j\textsuperscript{th} pricing slots, respectively. Similarly, we denote the respective
unit price values as $up_{i(init)}$, $up_{j(init)}$, $up_{i(opt)}$,
and $up_{j(opt)}$. We denote the $Total\thinspace cost$ as the sum of the product between aggregated demand and its corresponding unit price (\ref{eq:totalbill}) at each time interval. The difference between the optimised ($os_{h,i}$) and the preferred start time ($ps_{h,i}$) of each appliance multiplied by its penalty factor ($pf_{h,i}$) produces the inconvenience for the user of a given schedule. The sum of all the inconvenience produces the $Total\thinspace penalty$ for the community (\ref{eq:totalpenalty}), where $H$ is the total number of houses and $A_h$ is the number of appliances in the $h^{th}$ house. The sum of the $Total\thinspace bill$ and the $Total\thinspace penalty$ gives the $Total\thinspace cost$ (\ref{eq:totalcost}).

\begin{equation}
Total\,bill=\sum_{p=1}^{P}ad_{p}\thinspace\times\thinspace up_{p}\label{eq:totalbill}
\end{equation}

\begin{equation}
Total\,penalty=\sum_{h=1}^{H}\sum_{i=1}^{A_{h}}\mid os_{h,i}-ps_{h,i}\mid\thinspace\times\thinspace pf_{h,i}\label{eq:totalpenalty}
\end{equation}

\begin{equation}
Total\thinspace cost=Total\thinspace bill+Total\thinspace penalty\label{eq:totalcost}
\end{equation}

Before optimisation, all appliances are scheduled to run at their preferred start times. Thus, the $Total\thinspace cost$ is equal to the $Total\thinspace bill$, as no penalty needs to be applied. The objective of the optimisation is to minimise the $Total\thinspace cost$. However, the total demand remains unchanged (\ref{eq:fixeddemandinsystem}) as the DR scheme only moves devices to different slots without curtailing any demand. 

\begin{equation}
ad_{i(init)}+ad_{j(init)}=ad_{i(opt)}+ad_{j(opt)}\label{eq:fixeddemandinsystem}
\end{equation}

The optimisation calculates the $ad_{i(opt)}$ and $ad_{j(opt)}$ such that $Total\thinspace cost$ is minimised after the optimisation. We assume that the optimisation finds the values for $ad_{i(opt)}$ and $ad_{j(opt)}$ by moving $\alpha:(\alpha>0)$ amount of demand from the j\textsuperscript{th} pricing slot to the i\textsuperscript{th} pricing slot. Here, we assume that there is sufficient demand flexibility among the community and the penalty for flexibility is smaller than the benefit in unit price. Otherwise, the overall costs cannot be used as the objective function, as high penalties would remove the incentive to re-schedule demand. Note that these assumptions align with primitives of DR schemes that make a difference in the efficiency of energy distribution. Then, 
\[
ad_{j(opt)}=ad_{j(init)}-\alpha
\]
\[
ad_{i(opt)}=ad_{i(init)}+\alpha\thinspace\thinspace\thinspace\thinspace\because(\ref{eq:fixeddemandinsystem})
\]
\[
\therefore\thinspace\thinspace ad_{i(opt)}>ad_{i(init)\thinspace\thinspace\thinspace\thinspace}and\thinspace\thinspace\thinspace\thinspace ad_{j(opt)}<ad_{j(init)}
\]

And, 
\[
 up{}_{i(opt)}>up{}_{i(init)\thinspace\thinspace\thinspace\thinspace}and\thinspace\thinspace\thinspace\thinspace up{}_{j(opt)}<up_{j(init)}
\]
as the unit price monotonically increase with demand.

The $Total\thinspace penalty$ increases with the demand change. Considering the cost change,

\begin{multline*}
\Delta Total\thinspace cost = \\
ad_{i(init)}\times up_{i(init)}+ad_{j(init)}\times up_{j(init)}\\-
(ad_{i(opt)}\times up_{i(opt)}+ad_{j(opt)}\times up_{j(opt)}\\
+Total\thinspace penalty_{(opt)})
\end{multline*}

\begin{multline*}
\Rightarrow ad_{i(init)}\times up_{i(init)}+ad_{j(init)}\times up_{j(init)}\\-
((ad_{i(init)}+\alpha)\times up_{i(opt)}+(ad_{j(init)}-\alpha)\times up_{j(opt)}\\+Total\thinspace penalty_{(opt)})
\end{multline*}

\begin{multline*}
\Rightarrow ad_{i(init)}\times(up_{i(init)}-up_{i(opt)})\\
+ad_{j(init)}\times(up_{j(init)}-up_{j(opt)})\\
+\alpha\times(up_{j(opt)}-up_{i(opt)})-Total\thinspace penalty_{(opt)}
\end{multline*}

As we assume that the optimisation produces optimal results, $\Delta Total\thinspace cost > 0$. Thus, 
\begin{multline*}
ad_{i(init)}\times(up_{i(init)}-up_{i(opt)})
\\+ad_{j(init)}\times(up_{j(init)}-up_{j(opt)})\\
+\alpha\times(up_{j(opt)}-up_{i(opt)})-Total\thinspace penalty_{(opt)} > 0
\end{multline*}

In general, the optimisation finds an optimised demand distribution for any given $P$ pricing slots.
\begin{multline*}
{\displaystyle \sum_{p=1}^{P}ad_{p(init)}\times up_{p(init)}}-({\displaystyle \sum_{p=1}^{P}ad_{p(opt)}\times up_{p(opt)}+Total\thinspace penalty})>0
\end{multline*}
s.t
\[
{\displaystyle \sum_{p=1}^{P}ad_{p(init)}}={\displaystyle \sum_{p=1}^{P}ad_{p(opt)}}
\]

For the discussion of FDIAs, we assume a perfect optimisation oracle that produces the optimal demand distribution given the initial demand distribution and the set of penalty costs.

\subsection{False Data Injection Attack (FDIA)}
The commonly adopted DR architecture is vulnerable to FDIAs as utility companies have no control over data sources outside their perimeter~\cite{Dayaratne2019,Pilz2019}. Furthermore, HEMS networks utilise home area networks that are comparatively less secure and open up more injection opportunities for competent adversaries. In particular, this work considers an FDIA scenario, where initial demand forecasts are injected with false demand to gain financial benefits as an insider attack. Nevertheless, an external entity can also be the attacker given that they have sufficient attack motivations, including making customers dissatisfied with the existing provider or simply for the enthusiasm of hacking/compromising a system.
Moreover, adversaries do not require any tampering of communication channels, as they focus on manipulating HEMSs or devices that reside within their premises or in the premises of other participants. Possible attack strategies include 
\begin{enumerate}
\item Increase demand values of devices in the HEMS 
\item Add fake/real devices to the HEMS and remove them once the system converges
\item Collaborate with a set of other consumers or attack a set of consumers and distribute the false demand partially among them 
\end{enumerate}
Further, competent attackers can manipulate appliances' flexibility windows (earliest start and latest finish) on compromised HEMSs to create an artificial peak \cite{dayaratne2021we}. Such attacks will not alter the total demand in the system, as the fake demand borrowed from other timeslots. Recent work has shown that an adversary could gain up to 50\% cost reduction by injecting false demand into similar schemes~\cite{Dayaratne2019,Pilz2019,Weldehawaryat2017}. Even small cost reductions can collectively produce a significant gain for adversaries over time. 

We used different attacks including pulse, scaling, ramping, and random attacks. Attack formation is defined according to the type of demand injection~\cite{Cui2019}. 
In formal terms, given the initial forecasted aggregated demand for the i\textsuperscript{th} timeslot $ad_{i(init)}$, and the injected false demand ($fd_i$) the demand value utility company receives as the aggregated initial demand for the i\textsuperscript{th} slot is $ad_{i(init)}^{receive}$
\[
ad_{i(init)}^{receive} = ad_{i(init)} + fd_i
\]

Higher initial demand incurs a higher unit price for the attacked timeslot as the utility company requires more resources/reserves to generate the required amount. The resulting increased unit price triggers a set of HEMSs, given their flexibility, to reschedule their demand from the attacked time to other cheaper timeslots~\cite{Dayaratne2019,dayaratne2021we,Pilz2019}. However, since billing is dependent on real consumption and false demand is not actually consumed, the unit price used for billing purposes in the attacked timeslot is lower compared to the normal scenario~\cite{Dayaratne2019}. Therefore, the attacker (and anyone consuming energy during the same timeslot) pays a cheaper unit price for the slot.

\section{Resilient Device Scheduling}
The realisation of reliable and effective DR schemes heavily depends on proper security analysis of overall DR flows. Standalone optimisations cannot produce highly robust practical implementations while preserving strong privacy. Thus, anomaly detection and impact mitigation methods need to be added to design more robust DR schemes. One possible approach is to add appropriate detection and corrections on compromised forecasts before executing the DR optimisation.

\begin{figure*}[ht!]
    \centering
    \includegraphics[width=\linewidth]{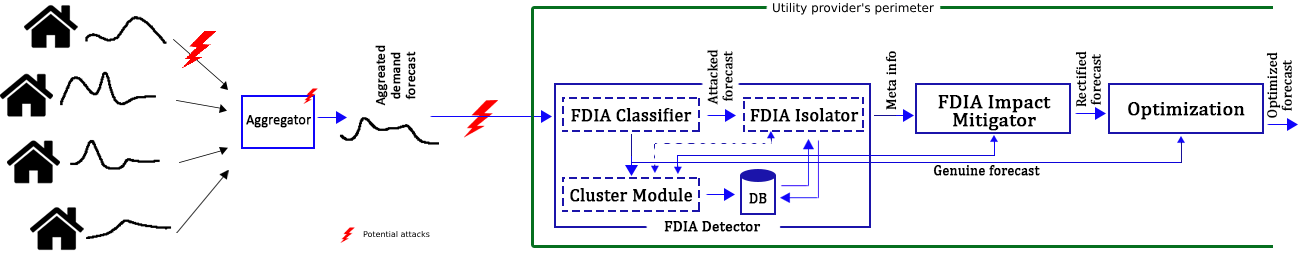}
    \caption{Resilient Device Scheduling Framework, potential attack points \& utility perimeter}
    \label{fig:frameworkd}
\end{figure*}

We leverage the theoretical framework proposed by Dayaratne et al.~\cite{Dayaratne2020a} shown in Figure \ref{fig:frameworkd}. The selection is motivated by the flexibility allowed by the framework to create an architecture where the implementer can decide on the optimal implementation without restrictions. It allows this flexibility by enabling the independent replacement of internal implementations within each module. The input to the framework is the aggregated community forecast, and the output is either the forecast itself or a corrected forecast. The framework comprises two main modules: the FDIA Detector and the FDIA Impact Mitigator. The Detector validates forecasts to identify any compromises, while legitimate forecasts are passed to the optimisation process. Compromised forecasts are handled by the Mitigator, which applies appropriate rectifications.

\subsection{FDIA Detector}
The Detector validates and produces meta-information on demand forecasts. Meta-information and the forecast itself are the outputs of the Detector. The Detector consists of FDIA Classifier, Cluster module, FDIA Isolator and a database.

\subsubsection{FDIA Classifier}
The FDIA Classifier determines whether a given demand forecast $df_i^D$ is manipulated with FDIAs, where $D$ is either 24 (hourly)/48 (half-hour) or any other value depending on the DR scheme. Forecasts that are classified as attacked are routed to the Isolator. Genuine forecasts are forwarded to the optimisation and Cluster module. FDIA detection is a part of the mainstream work of anomaly detection. Unsupervised and supervised classification or regression methods can be used to implement the Classifier. In complex DR optimisations, additional inputs, such as weather patterns and demographic information, need to be included in the detection as they can potentially be manipulated and lead to failures in forecasts. Thus, detection features should be carefully chosen. Ensemble methods can enhance the detection, given the availability of sufficient time and computing resources. The Classifier needs to ensure a high detection rate with a minimum false-positive rate as false positives can lead to additional costs amidst manual inspections. 

\subsubsection{Cluster Module}
\label{cluster_module}
Attack-free forecasts (forwarded by the Classifier) are the input to the Cluster module. The module groups the inputs into $k$ clusters, where $k$ can be determined empirically to capture prevalent aggregated forecasts. A formed cluster represents a generalised demand pattern of the connected community. The number of clusters ($k$) can be adaptive or fixed depending on factors, such as seasonality and a range of stored data. Cluster centroids and forecasts with their corresponding cluster indices are stored in the database. k-medoids, k-means, or sequential k-means algorithms are suited for the Cluster module, given their low complexity compared to other clustering approaches.

\subsubsection{FDIA Isolator}
Attacked slot/values(s) of a given attacked forecast must be correctly determined to implement corrections. In formal terms, given an attacked demand forecast ($df_i^D$), the Isolator determines the manipulated timeslots/values. Outlier aspect mining techniques, such as Isolation Path Score, Local Outlier Factor (LOF), Density-Based z-Score and Spectral Residual \cite{Vinh2016,Ren2019} can be used to implement the Isolator. The Isolator first determines the closest cluster of the attacked forecast using the Cluster module. The identified cluster forecast and the attacked forecast are used to identify the attacked slots using an outlying aspect mining method.

\subsection{FDIA Impact Mitigator}
The Mitigator determines and executes adequate impact mitigation strategies, given attacked forecasts and attacked slots/values. An effective mitigation strategy must ensure the following. 
\begin{enumerate}
    \item Adversary's gain should be minimised
    \item Other users should not be negatively impacted by the rectification
\end{enumerate}

Attack prevention is always better than mitigation. However, prevention is arduous, as HEMSs reside outside the utility companies' control perimeter. Further, common security mechanisms such as encryption, sealed devices, authentication, authorisation have little to no impact on preventing FDIAs. Thus, utilities need to detect anomalies and then minimise or eliminate the potential impact. 
The strategy chosen in our approach is to correct the potential false data to the closest values of the corresponding attack-free scenarios. Some possible strategies to achieve this are discussed below.

\subsubsection{Threshold based filters}
Utility companies can use predefined thresholds to curtail false data over a certain threshold by leveraging previous demand patterns or locally generated patterns. Reliable results can be achieved when the generated/previous patterns are closely aligned with the forecast. Nevertheless, uncertainties in exogenous inputs and possible manipulation in load forecasting models can produce inaccurate forecasts. 

\subsubsection{Replaceable forecasts}
In this approach, the entire forecast or attacked values are replaced with locally computed or mapped historic forecast (e.g. previous day, corresponding day of the earlier week). However, consumers can be negatively affected if the replaced values are significantly different from the actual values. Further, a forecast can vary from its closest forecasts as consumption are inherently chaotic. 

Commonly adopted time-series forecasting methods such as AR, MA, ARMA, ARIMA, TBATS, prophet model or complex ML-based forecasting methods can be used with both threshold and replaceable methods. 

\section{Implementation \& Experiment}
\subsection{FDIA Detector}
\subsubsection{FDIA Classifier Implementation}
FDIA detection can be perceived as a time-series classification problem. Thus, we used the convolutional neural network (CNN) model proposed in the state of the art deep-learning-based time-series classification literature \cite{zhao2017convolutional,fawaz2019deep}. The model is composed of five layers, and the input layer has 48 neurons. The first and the third layers are convolution layers, and each is followed by an average pooling layer. Convolution layers consist of 6 and 12 filters, respectively, where the kernel size is seven. The output layer is a fully-connected layer with two neurons. The sigmoid function was used as the activation in all layers, while the mean squared error (MSE) was used as the loss function. We use these parameters as suggested for both univariate and multivariate time-series classification \cite{zhao2017convolutional}.

In parallel, we have also used recently proposed unsupervised CSR (Cluster based Spectral Residual) \cite{dayaratne2022fdia}, which integrates both the k-means and the spectral residual methods. The combined unsupervised model has two hyper-parameters (the value of $k$ and $q$ except of the threshold) to be configured. Parameter values can be set empirically. The pseudocode of the CSR model is shown in Algorithm \ref{algo:k-means-sr}. Several clusters of attack-free demand forecasts were formed using the k-means clustering, as explained in section \ref{cluster_module}. Then, squared Euclidean distance (Equation \ref{eq:kmeans-distance}) is used to determine the closest cluster centroid among established clusters for each incoming forecast (line 3). The corresponding cluster centroid is subtracted from the incoming forecasts to retrieve the residual forecasts (line 4). Saliency maps for residual forecasts are calculated using SR (lines 5-10). A demand forecast is classified as attacked if the maximum value of the saliency map obtained is higher than a threshold. We have used maximum values from saliency maps as maximum values corresponding to the anomalous time-slots.	
\begin{algorithm}	
\SetAlgoLined 	
\KwIn{A set of $N$ demand forecast vectors representing half hour demand prediction $\{df_1^D, df_2^D, .., df_N^D\}$ (containing both train and test sets), k, threshold} 	
\KwOut{attack/normal labels for demand forecast on test set} 	
Apply k-means clustering as explained in Section~\ref{cluster_module} on filtered train test with $k$ cluster centroids $\mu_j^D$ $(1\leq j \leq k)$ \\	
\For{each demand forecast $df_i^D$ in test set}{     	
  find closest cluster centroid $\mu_j^D$\\	
  Get the residual $rs_i^D \gets df_i^D - \mu_j^D$\\ 	
  amplitude spectrum $as_i \gets ||\mbox{\it fast fourier transform}(\mathit{rs_i^D)}||$\\	
  phase spectrum $ps_i \gets \mbox{\it phase(fast fourier transform}(\mathit{rs_i^D}))$\\ 	
  log spectrum $ls_i \gets \log{(\mathit{as_i})}$\\ 	
  avg log spectrum $als_i \gets \mbox{\it convolute}(\mathit{as_i}))$\\	
  Spectral Residual $rs_i \gets als_i-ls_i$\\ 	
  saliency map $sm_i^D \gets ||\mbox{\it inverse fast fourier transform}(\exp(\mathit{rs}_i+\mathit{ips}_i))||$

  \eIf{$\max(sm_i^D) \geq threshold$}
  {
      $df_i^D$ is an attacked forecast\\
      $attacked index \gets index(\max(sm_i^D))$\\
  }{
      $df_i^D$ is a normal forecast\mbox{}
  }
}

\caption{Pseudocode of the CSR model}
\label{algo:k-means-sr}
\end{algorithm}

\subsubsection{Cluster Module Implementation}
Anomaly detection and mitigation for DR optimisation must be carried out during the actual scheduling process. Therefore, we used k-means clustering to implement the Cluster module with low complexity and low computation compared to other clustering approaches. k-means clustering groups $N$ objects into $k$ groups, where objects in the same group have shorter distances between members and to the centre of the group than to other members in the dataset. In our context, demand forecasts ($df_{1},df_{2},..,df_{n}$ where each forecast is a $D$-dimension vector), are grouped into $k$ partitions ($T_{1},T_{2},..,T_{k}$) by minimising the sum of squares between clusters using (\ref{eq:kmeans}), where $\mu_{i}$ is the mean demand forecast (centroid) in $T_{i}$.

\begin{equation}
{\displaystyle {\textstyle \underset{\mathrm{T}\,}{arg\,min}\sum_{i=1}^{k}\sum_{df_{i}\epsilon\mathrm{T}}\mid df_{i}-\mu_{i}\mid^{2}}}\label{eq:kmeans}
\end{equation}

The closest cluster for given forecast $df_{i}$ is the cluster $T_{j}$ that minimises the squared Euclidean distance (\ref{eq:kmeans-distance}) between $df_{i}$ and $\mu_{j}$.

\begin{equation}
min(\parallel df_{i}-\mu_{j}\parallel^2)\label{eq:kmeans-distance}
\end{equation}
We have used both Euclidean and Dynamic Time Warping (DTW) distances in our experiments implemented using the \textit{tslearn} python package. As DTW did not provide significant improvements over Euclidean distance, Euclidean distance was selected to minimise the computational overhead.

\subsubsection{FDIA Isolator Implementation}
The Isolator uses the cluster module to calculate the closest cluster index of the received attacked forecast and identifies anomaly time slots of the detected forecast. Closest attack-free forecasts from the database are used to detect the attacked time slots. 

We have adopted the Beam search based isolation path score method~\cite{Vinh2016} to implement the Isolator. It uses the isolation forest method with some alterations. The isolation forest method leverages the fact that outliers are isolated as they are few and different. Hence, outliers (attacked time-slots in our use-case) reside closer to the root and have shorter path lengths when employing random binary trees as the isolation mechanism. The Isolation path score method provides the required computational efficiency in terms of real-time processing and scalability as the underlying isolation forest does not rely on distance computation. The method isolates the anomaly from a dataset using a series of random binary splits. Each split divides the data space into two halves. The splitting part ends when the anomaly point is isolated or until all the remaining data points have the same value for the randomly selected feature that been used. When the splitting ends with more than one point, the path length is increased by $2(ln\mid X\mid+\gamma)-2$, where $\gamma=0.5772$ is the Euler constant. Multiple sub-samples of the data set is used in an ensemble approach to obtain the path length and the averaged path score is used for the final calculation. The Beam search framework initially explores 1-dimension features (individual demand values of each time slot) to detect apparent attacks. All 2-dimension sub-spaces (combinations of time-slot pairs) are explored in the second stage. Each of the top $l$ attack-positive subspace from the second stage appends with another feature (time-slot) in the next phase and executes the search. This process iterates up to a pre-defined limit of sub-spaces. We limit the maximum sub-spaces to three, assuming an attack will not last more than 1.5 hours. The isolation path score was calculated for all the explored sub-spaces in each stage. The highest scored sub-spaces are detected as attacked slots. 

We benchmarked the beam search against the LOF method, which can be used in outlying aspect detection tasks such as in our use-case determining the attacked slots. Leveraging the deviation of a given data point concerning its neighbours, the LOF can determine anomalous points. In particular, LOF uses the distance of k-nearest neighbours of a data point in comparing the local density of data points \cite{breunig2000lof}.
In formal terms, given a data point $X$ in set $D$, we can define the k-distance of the $X$ as $dist_k(X)$, which is the distance between $X$ and its' k\textsuperscript{th} nearest neighbour. For example, if $k=3$, then the k-distance of $X$ is the distance between $X$ and its' third nearest neighbour. Using the k-distance, we can define the k-distance neighbourhood of $X$ ($N_k(X)$) as all the data points that reside in an area where the distance from $X$ to any point is not more than $dist_k(X)$ (\ref{eq:lof_k_neighbourhood}).
\begin{equation}
N_k(x)=\{q \epsilon \, D|dist(q,X)\leq dist_{k}(X)\}
\label{eq:lof_k_neighbourhood}
\end{equation}
Then, the reachability distance of $X$ can be defined as the maximum of the distance of two points and the k-distance of the second point (\ref{eq:lof_reachability_distance}).
\begin{equation}
reach{-}dist_k(X,q)=max(dist_k(X), dist(X,q))
\label{eq:lof_reachability_distance}
\end{equation}
The reachability distance is then used to calculate the local reachability density (LRD). LRD is the inverse of the average of reachability distance of k nearest neighbours of $X$ (\ref{eq:lof_local_density}).
\begin{equation}
LRD_{X}=\frac{1}{\frac{\sum_{x'\epsilon N_{k}(X)}reach{-}dist_{k}(X,x')}{\parallel N_{k}(X)\parallel}}
\label{eq:lof_local_density}
\end{equation}
Finally, LOF of $X$ can be calculated as the average of the ratio of local reachability of $X$ and its' k neighbours (\ref{eq:lof_lof}). LOF of a point is the average ratio of the LRDs of the neighbours of the point and the LRD of the point.
\begin{equation}
LOF_{k}(X)=\frac{\sum_{x'\epsilon N_{k}(X)\frac{LRD_{k}(x')}{LRD_{k}(X)}}}{\parallel N_{k}(X)\parallel}\label{eq:lof_lof}
\end{equation}

If the LOF value is significantly greater than 1, that point can be considered as an attacked time slot.

The CSR model that we introduced in the classifier section was used as the second bemchmark to detect the attacked time slots. The underlying idea is that the max value index of the obtained saliency map corresponds to the attacked slot(s). The corresponding code is depicted in line 13 of Algorithm \ref{algo:k-means-sr}.

\subsection{FDIA Impact Mitigator Implementation}
The goal of the Mitigator is to compute a variant of the aggregated forecast that removes false data while keeping genuine data as unchanged as possible. We analysed six different implementations for the Mitigator as follows. 

\begin{itemize}
    \item \textbf{Method 1} - Uses the clustering to find the closest legitimate pattern and then replaces the demand values of the attacked time slots (of the attacked forecast) by the corresponding demand values from the legitimate pattern in each iteration
    \item \textbf{Method 2} - Uses the clustering to find the closest legitimate pattern and then replace the demand values of the attacked time slots (of the attacked forecast) by corresponding demand values from the legitimate pattern in the initial forecast. Starting from the second iteration, calculate the change of demand for the attacked time slots between current and previous iteration. Add/subtract the amount of demand change to/from the corresponding demand values of the legitimate pattern and replace the attacked values with the newly calculated values. Finally, assign the corrected demand pattern as the legitimate pattern for the next iteration.
    \item \textbf{Method 3} - Uses the clustering to find the closest legitimate pattern and then replaces the demand values of the attacked time slots (of the attacked forecast) by the corresponding demand values from the legitimate pattern. Then use the clustering again to find the closest legitimate pattern for the corrected forecast. Replace the demand values of the attacked time slots (of the corrected forecast) by the corresponding demand values of the newly identified legitimate pattern. Starting from the second iteration, use the newly identified legitimate pattern as the legitimate pattern for all other iterations and replace the attacked values by the corresponding demand values from the legitimate pattern.
    \item \textbf{Method 4} - This is a combination of Methods 2 and 3. Repeat the Method 3 during the initial forecast. Starting from the second iteration, incorporate the change of demand between current and previous iteration for the attacked slots as in Method 2 with the legitimate pattern. 
    \item \textbf{Method 5} - Uses interpolation to find the closest legitimate pattern. In particular, it uses forward linear interpolation (using the values up to the one slot prior to the attacked slot) and interpolates the demand values for the attacked slots. Corresponding demand values from the interpolated forecast are used to replace the demand values of the attacked time slots in each iteration as in Method 1.
    \item \textbf{Method 6} - This is a combination of Methods 2 and 5. Repeat Method 5 during the initial forecast. Starting from the second iteration, incorporate the change of demand between current and previous iteration for the attacked slots as in Method 2 with the legitimate pattern (interpolated forecast). 
\end{itemize}

Each implementation starts by retrieving attacked time slots from the Detector. Next, for each attacked forecast, the closest legitimate pattern (an attack-free forecast that closely aligns with the attacked forecast for all the time slots except for the attacked slots) was determined. However, with double clustering, finding the closest attack-free forecast is two-fold. Firstly, the attacked slots are replaced with the initially identified closest forecast. The corrected forecast is then (re)used to find the closest forecast repeating the previous step. In interpolation methods, forward linear interpolation (using the values up to attacked slot) is used to calculate the values of attacked slots, and the calculated forecast is used as the closest attack-free forecast. Each method replaced attack values with corresponding legitimate pattern values. In Algorithm \ref{algo:mitigator}, lines 1 to 20 depict the pseudocode of the initial correction process.  

\begin{algorithm}
\SetAlgoLined 
\KwIn{An attacked initial demand forecast $df^D$ representing half-hourly demand prediction $\{df_1^D, df_2^D, .., df_{48}^D\}$ and attacked indices} 
\KwOut{Rectified demand forecast $rdf^D$ representing half-hourly corrected demand prediction $\{rdf_1^D, rdf_2^D, .., rdf_{48}^D\}$} 

\eIf{Initial}{
    \tcp{initial forecast (during 1st iteration)}
    \eIf{Single cluster or Double cluster}
    {
        $cp^D \gets$ closest legitimate cluster pattern of $df^D$
    }
    {
        \For{each index $i$ in attacked slots}{ 
            $df_i^D \gets linear\thinspace interpolation(df_1^D,.. df_{i-1}^D)$\\
        }
        $cp^D \gets interpolated\thinspace forecast$\\
    }
    $rdf^D \gets df^D$\\
    
    \For{each index $i$ in attacked slots}{
        $rdf_i^D \gets cp_i^D$
    }
    \If{Double cluster}{

        Find the closest legitimate cluster pattern $cp^{D^`}$ of $rdf^{D}$
        \For{each index $i$ in attacked slots}{
            $rdf_i^D \gets cp_i^{D^`}$
        }
        $cp^D \gets cp^{D^`}$\\
    }
}
{
\tcp{during all other iterations}
\For{each index $i$ in attacked slots}{
    \eIf{Fixed}{
        $rdf_i^D \gets cp_i^D$\\
    }
    {
        $\delta_i \gets df'_i - df_i$ \tcp{difference between the previous and current iteration forecast for the slot}
        $rdf_i^D \gets cp_i^D - \delta_i$\\
    }
}
\If{Adaptive}{
    $cp^D \gets rdf^D$\\
}
}
Sends $rdf^D$ to the optimisation\\
\caption{\label{algo:mitigator}Pseudocode of the correction algorithm}
\end{algorithm}

\begin{table*}
\centering
\caption{Different implementations of The Mitigator\label{tab:mitigator_impls}}
\begin{tabular}{ cccccc }\hline 
 & \multicolumn{3}{c}{\textbf{Based on}} & \multicolumn{2}{|c}{\textbf{Rectification}}\tabularnewline
\cline{2-6} 
& \textbf{Single cluster} & \textbf{Double cluster} & \textbf{Interpolation} & \textbf{Fixed} & \textbf{Adaptive}\tabularnewline
\midrule
Method 1 &  \checkmark &  &  & \checkmark & \tabularnewline
Method 2 &  \checkmark &  &  &  & \checkmark\tabularnewline
Method 3 &  & \checkmark &  & \checkmark & \tabularnewline
Method 4 &  & \checkmark &  &  & \checkmark\tabularnewline
Method 5 &  &  & \checkmark & \checkmark & \tabularnewline
Method 6 &  &  & \checkmark &  & \checkmark\tabularnewline
\bottomrule 
\end{tabular}
\end{table*}

Given that the optimisation utilises multiple iterations to converge and the source of the attack cannot be eliminated, rectifications need to be performed in each iteration. The pseudocode for the correction during iterations is depicted in Algorithm \ref{algo:mitigator}, from lines 21 to 32. The demand values of the attacked indices (after the initial iteration) are always replaced with the corresponding values from the closest legitimate pattern identified previously for the methods (Methods 1, 3, and 5) that use fixed corrections (lines 22-24). Retaining the demand values fixed for the attacked slots over the iterations, the optimisation algorithm prevents attempting to optimise the underlying genuine forecast for those time slots under fixed methods. The adaptive approach is used to solve this problem. The adaptive methods (Methods 2, 4, and 6) also replace the attacked values (on the initial forecast) similar to the fixed methods. However, during iterations, the demand values of the attacked time slots are adjusted to allow the possible optimisation of the genuine demand in the attacked slots, considering the difference between the previous forecast ($df_i^{'}$) and the current forecast ($df_i$) for the attacked slots (lines 26-27). The corrected forecast is then used as the closest legitimate pattern for the next iteration. Table~\ref{tab:mitigator_impls} summarises the correction methods.

\subsection{DR Implementation}
We used the distributed device scheduling algorithm ~\cite{He2018} introduced by He et al. to analyse the impacts of FDIAs, as it aligns with the abstract model that is explained in Section~\ref{sec:prel}. The distributed optimisation is highly efficient and scalable with strong privacy due to limited information exchange between households and the utility company. 
However, it should be noted that neither the attack nor the combination of detection and mitigation is specific to this particular algorithm. 

\subsection{Dataset}
We extracted minute-wise energy consumption data for appliances from 168 houses in Austin, Texas, for a three-month period between June and August 2017 from the Pecan Street data portal~\cite{Pecanstreet}. The consumption data was then mapped to the appliances' demand, as the original data did not include demand values. Following a similar approach mentioned in \cite{Dayaratne2019,He2018}, the daily demand profiles of individual households were aggregated to generate 92 demand forecasts, each representing half-hour demand. To create more normal demand profiles, we modified the preferred start times of a randomly selected number of appliances from randomly selected households each day. Additionally, we injected false demand ranging from 0.1\% to 25\% of the daily controllable demand (resulting in an overall demand increase between 0.01\% - 2.5\%) at different time slots, producing a set of attacked profiles. The dataset consists of a total of 277,752 forecasts, with approximately 5\% of the forecasts being attacked profiles.
It is important to note that we used aggregated demand forecasts instead of individual households' forecasts to address privacy concerns. The dataset was divided into train and test sets, with the training dataset comprising two-thirds of the data. The test dataset includes 91,659 demand forecasts, of which 4,946 are attacked forecasts, representing approximately 5\% of the test data.

\section{Evaluation}
\subsection{FDIA Detector}
Two of the recently proposed state of the art supervised and unsupervised anomaly detection models were used to benchmark the classifier. The supervised model is an FDIA detection method that uses the combination of k-means clustering with a Naive-Bayes classifier (kNBC) \cite{Cui2019}. The unsupervised model uses the spectral residual (SR) in time-series anomaly detection \cite{Ren2019a}. Accuracy, Precision, Recall, F1 and False positive rate (FPR) were used in the evaluation. The proposed CNN model outperforms the benchmarked models as depicted in Table \ref{tab:Result-comparison-between}, where it produces the best accuracy, precision, F1 and FPR. Unsupervised CSR model produces the best recall among all the models.



\begin{table}[!b]
\centering
\caption{Evaluation of FDIA Classifier\label{tab:Result-comparison-between}}
\begin{tabular}{lccccc}
\toprule
Method & Accuracy & Precision & Recall & F1 & FPR\tabularnewline
\midrule
SR\cite{Ren2019a} & 90.38\% & 13.99\% & 15.20\% & 14.57\% &5.33\%\tabularnewline
kNBC \cite{Cui2019} & 93.58\% & 41.49\% & 46.36\% & 43.79\% & 3.73\%\tabularnewline
CSR & 95.39\% & 54.61\% & \textbf{86.53\%} & 66.96\% & 4.10\%\tabularnewline	
CNN & \textbf{97.54\%} & \textbf{94.42\%} & 57.84\% & \textbf{71.74\%} & \textbf{0.19\%}\tabularnewline\bottomrule
\end{tabular}
\end{table}

\subsection{FDIA Isolator}
\begin{figure}	
\centering	
\includegraphics[width=\linewidth]{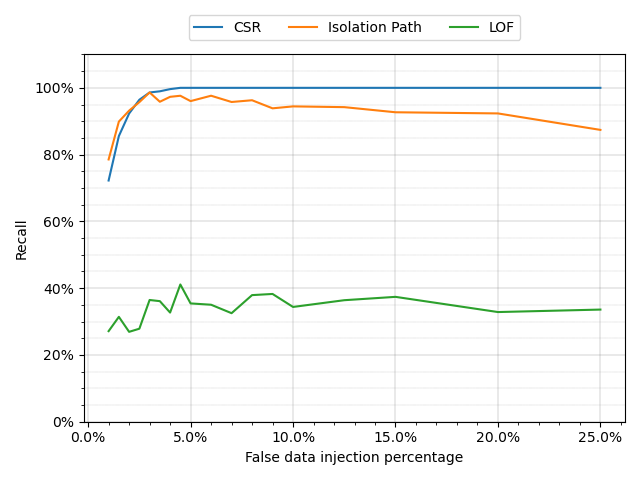}	
\caption{Variation in Recall with different injection percentages}	
\label{fig:recall_percentage_slot}	
\end{figure}

A filtered dataset that only comprises attack-free forecast was used to train the Cluster module with the $k$ value is set to 400. The cluster indices of individual forecasts were saved in the database. Attacked forecast and corresponding attack-free forecasts of the same cluster (retrieved from the database) are used with the Beam search to determine attacked time slots. The LOF method is used to benchmark the performance of the Beam search. LOF was selected in the benchmark as it is one of the commonly used outlying aspect mining methods. The Isolator correctly detected the precise time-slots of 4631 attacked forecasts out of 4946, which results in 93.63\% detection accuracy. In contrast, the LOF method did not produce a significant result as its recall was 34.05\%. Both the CSR and Isolation Path methods produce recall values above 90\%, which is highly desirable given the requirement. In particular, the CSR model increases the recall value by 3.25\% compared to the Isolation paths method. The Isolation path method correctly detected the precise time-slots of 4631 attacked forecasts out of 4946, which results in 93.63\% detection accuracy. In contrast, the LOF method did not produce a significant result as its recall was 34.05\%. 	
Table \ref{tab:slot_diff_percentage} shows the correctly detected attacked slots per each distinct injection percentage. The variation in the recall value for detected attacked slots with each injection percentage is depicted in Figure \ref{fig:recall_percentage_slot}. The CSR model's recall values was only 72\% for the lowest injection. Nevertheless, the CSR model's recall value increases with injection percentage and reaches 100\% for percentages over 4\%. However, a similar variation cannot be observed in other models. Nevertheless, the Isolation path method performs better compared to the CSR model with lower injection percentages. 	
\begin{table}	
\centering	
\caption{Recall values of attacked time slot detection}	
\begin{tabular}{lc}	
\toprule 	
Method & Recall\tabularnewline	
\midrule	
LOF & 34.05\%\tabularnewline	
Isolation Path & 93.63\%\tabularnewline	
CSR & \textbf{96.68\%}\tabularnewline	
\bottomrule 	
\end{tabular}	
\label{tab:time_slot_detect_result}	
\end{table}	
\begin{table}	
\centering	
\caption{Correctly detected attacked slots per each injection percentage}	
\begin{tabular}{rcccr}	
\toprule 	
\multirow{2}{*}{Percentage} & \multirow{2}{*}{Attack count} & \multicolumn{3}{c}{Detected count}\tabularnewline	
\cline{3-5} 	
 &  & CSR & Isolation Path & LOF\tabularnewline	
\midrule 	
1\% & 317 & 229 & 249 & 86\tabularnewline	
1.5\% & 277 & 237 & 249 & 87\tabularnewline	
2\% & 234 & 216 & 218 & 63\tabularnewline	
2.5\% & 280 & 270 & 268 & 78\tabularnewline	
3\% & 288 & 284 & 284 & 105\tabularnewline	
3.5\% & 288 & 285 & 276 & 104\tabularnewline	
4\% & 260 & 259 & 253 & 85\tabularnewline	
4.5\% & 253 & 253 & 247 & 104\tabularnewline	
5\% & 302 & 302 & 290 & 107\tabularnewline	
6\% & 254 & 254 & 248 & 89\tabularnewline	
7\% & 283 & 283 & 271 & 92\tabularnewline	
8\% & 269 & 269 & 259 & 102\tabularnewline	
9\% & 277 & 277 & 260 & 106\tabularnewline	
10\% & 288 & 288 & 272 & 99\tabularnewline	
12.5\% & 294 & 294 & 277 & 107\tabularnewline	
15\% & 246 & 246 & 228 & 92\tabularnewline	
20\% & 274 & 274 & 253 & 90\tabularnewline	
25\% & 262 & 262 & 229 & 88\tabularnewline	
\hline	
Total & 4946 & 4782 & 4631 & 1684 \tabularnewline	
\bottomrule 	
\end{tabular}	
\label{tab:slot_diff_percentage}	
\end{table}

\subsection{FDIA Mitigator}
Appliance wise, half-hourly demand profiles of 92 days are used to evaluate the Mitigator. We configured a device to operate on a peak scheduling slot of each day for a pre-defined duration. The device bill and the community cost were calculated once the DR optimisation converged. These values depict the cost of the attacker and the cost of the community in the attack-free scenario. Subsequently, we inject fake demand into the initial demand distribution of the day and rerun the optimisation. The cost values are recalculated. The new values depict the results without any corrections. We repeat the same attack with the FDIA mitigator to calculate the cost values with corrections. The amount of fake demand was gradually increased from 0.1\% up to 5\% in different attack scenarios. Cost values were calculated in each scenario. The same steps were repeated for the 92 days to obtain 1656 attack-free, attack without correction, and attack with correction scenarios. 

\begin{figure*}
    \centering
        \includegraphics[width=\linewidth]{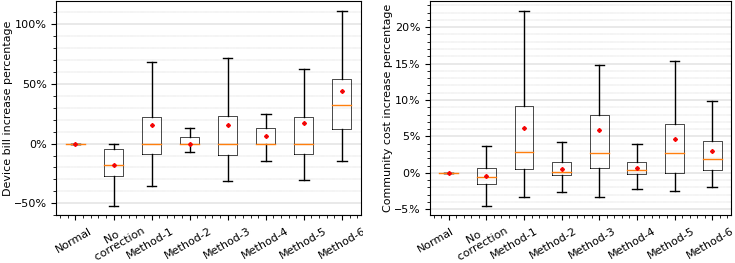}
    \caption{Attacker's device bill and community's total cost variation with different corrections}
    \label{fig:cost_variation}
\end{figure*}

The left side graph in Figure \ref{fig:cost_variation} depicts the variation of the attacker's device bill with different correction methods. The attacker consistently achieved a cost reduction without any correction. Mean and median values were -17.7\% and -17.9\%, respectively. However, the attacker could not obtain a substantial bill reduction with the correction methods. The mean and median values for each correction scenario were 0\% or greater than 0\%. The adaptive interpolation method increases the device bill by 43. 9\% on average, which is the highest increase among all methods, whereas the fixed interpolation method increases the device bill by 16.87\% on average. Fixed clustering methods result in a significant increase in device bills compared to adaptive methods. Adaptive clustering methods neither increase nor decrease device cost substantially. Fixed correction methods result in a significant increase in community cost compared to adaptive methods, as shown on the right side of Figure \ref{fig:cost_variation}. On the contrary, adaptive methods 2, 4, and 6 increased the community cost by 0.53\%, 0.67\% and 2.98\%, respectively. 

\begin{figure}
    \centering
    \includegraphics[width=\linewidth]{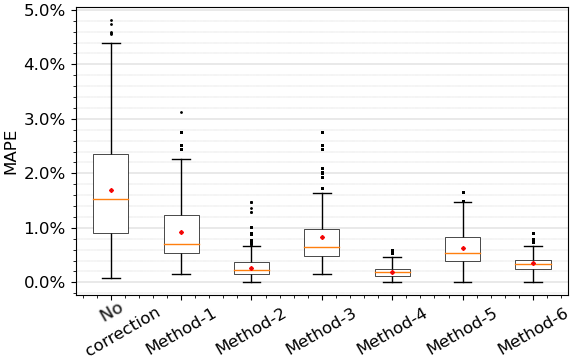}
    \caption{Variation in MAPE values of different correction methods}
    \label{fig:correction_mape}
\end{figure}

We also used the mean absolute percentage error (MAPE) to determine how close the rectified forecasts are to corresponding non-attack scenarios. Comparatively, adaptive methods have lower MAPE values, as shown in Figure \ref{fig:correction_mape}. Moreover, adaptive clustering methods produce corrected forecasts that are much closer to the corresponding non-attack scenarios. In particular, the MAPE value of 0.19\% of the adaptive double-clustering method is the minimum among all methods, where the other two adaptive methods' MAPE values are 0.26\% and 0.36\%

\section{Discussion}
Financial interests or other malicious intent can drive consumers and external attackers to execute FDIAs on DR systems. Thus, resilient implementations of distributed device scheduling are imperative to ensure the stability and fairness of schemes. 

The CNN-based classifier achieved the highest detection rate (excluding recall) while maintaining a minimised FPR. A low FPR ensures cost efficiency since manual interventions for inspection are significantly more expensive. It should be noted that this study did not consider instances of bad data or significantly small injections, as the optimisation process would not be significantly affected by a small amount of such data/injections to yield meaningful benefits.
However, the recall value of the classifier is noticeably low for attacks with small injection percentages. Specifically, the classifier only detected 1,538 out of 3,574 attacks where the injection percentage was between 0.01 and 0.09\%. Considering the potential impact of these attacks, further improvements are necessary.
Nevertheless, acquiring a sufficient amount of labeled data for training supervised models is a challenging task. Therefore, we believe it is more appropriate to explore and enhance the use of unsupervised methods in detecting FDIAs. In particular, the unsupervised CSR model achieved the highest recall rate. Although the Isolator demonstrates high detection accuracy, it fails to correctly identify the attacked slots in almost 3\% of the attacks involving manipulated forecasts. Incorrectly labeling regular time slots as attacked slots can have a negative impact on the community, as these detected time slots are used in the correction phase. Hence, improving the accuracy of the Isolator is essential to enhance the overall resilience of the system.

Distributed detection of FDIAs within HEMS can improve the detection accuracy. However, implementing comprehensive HEMS-level detection is impractical due to resource constraints and the potential for skilled adversaries to bypass or disable local detectors. It is also challenging for individual detectors to identify attacks involving small injections across multiple HEMSs, where aggregated small injections collectively create sufficient fake demand to deceive the DR system. Collaboration among local detectors offers a potential solution. However, determining shared data and selecting appropriate collaborators raises privacy concerns. Therefore, combining local detection within HEMSs and detection at the aggregator level is a more practical approach to enhance detection capabilities.

\begin{figure}
    \centering
    \includegraphics[width=\linewidth]{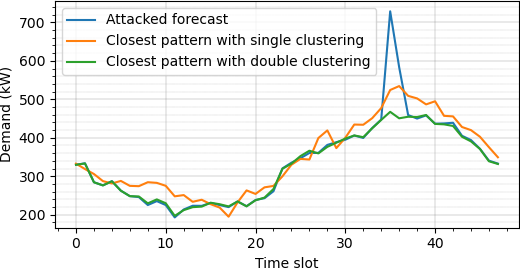}
    \caption{Single clustering vs Double clustering}
    \label{fig:double_cluster}
\end{figure}

Fixed correction methods can significantly increase the attacker's bill. However, the increase can negatively impact other households that have scheduled appliances at the same time slots, where total community cost significantly increased with fixed correction methods. Thus, adaptive correction methods are more suitable as they produce minimised cost increase for the community. Further, adaptive correction methods can significantly reduce the attacker's gain without negatively impacting other houses operating on the same time slots. The double clustering method can eliminate issues in distance measures when determining the closest cluster for a forecast that contains a significant false demand. As shown in Figure \ref{fig:double_cluster}, using a single clustering approach did not determine the closest pattern in some attacks. Thus, applying the double clustering approach is more effective in such scenarios. The rectified forecasts of double clustering methods are significantly adjacent to the corresponding non-attack forecasts. However, compared to the single clustering, there is an associated overhead in the computation of the double clustering approach. However, this overhead can be ignored, since more accurate forecasts are determined using the double clustering approach. Simple interpolation methods cannot capture the stochastic nature of demand. Thus, interpolation methods are less effective compared to clustering methods. 

\begin{figure}
    \centering
    \includegraphics[width=\linewidth]{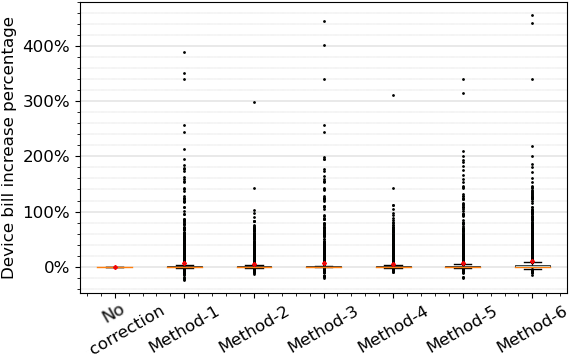}
    \caption{Device bill increase percentage under different correction methods when the demand is consumed}
    \label{fig:whatif_real}
\end{figure}

Recent work has highlighted that it is significantly difficult to predict accurate demand values with the existing load forecasting method \cite{Schreck2020}. Further, it is arduous to ensure the correctness of forecasts as attackers can manipulate exogenous parameters in load forecasting. Thus, we argue that historical patterns in corrections are more effective compared to computing forecasts locally. Given that the framework eliminates significantly high demands from the detected time slots, the demand concentration of time slots (time slots with the genuine high demand) will be higher than the no-correction scenario. Higher demand causes the unit price to be higher than the no-correction scenario, which increases the cost of consumers who genuinely forecast their high demand. Figure \ref{fig:whatif_real} depicts how a genuine consumer's device bill vary if they consume the high demand that they forecast.  In particular, this increase affects all the consumers who have some devices scheduled in those time slots. Thus, possible avenues to differentiate a genuine forecast with high demand from attacks require further experiments. 

Local mitigation involves reconstructing compromised demand forecasts at the HEMS level. However, individual house-level forecasts exhibit significant diversity due to various consumption-influencing factors and their constant evolution. Additionally, adversaries may bypass or disable local correction methods, posing challenges to individual forecast rectification. Therefore, a multi-step rectification strategy that combines local and aggregator-level approaches is more effective in addressing these challenges.

\section{Conclusion}
This work analysed and implemented a resilient device scheduling DR scheme against FDIAs. We used both theoretical and experimental results in the analysis. Existing anomaly detection methods and a novel impact mitigation method were integrated to implement a resilient framework. The implemented framework was evaluated using a real-world dataset. Though the framework produces notably high results, further improvements are required to enhance the resiliency of distributed device scheduling. In particular, more work is required on FDIA detection and identifying attacked slots, as incorrect detection results can negatively impact the community and utility companies. Additionally, proper analysis of load forecasting methods is necessary to combine local forecasting methods with correction.

 \bibliographystyle{elsarticle-num} 
 \bibliography{reference}





\end{document}